\documentclass[epj]{svjour}
\usepackage{graphics}
\usepackage{amsmath}     
\newcommand{\dir}{Figs}
\newcommand{\ie}{{\em i.e.}, }
\newcommand{\eg}{{\em e.g.}, }
\newcommand{\etal}{{\em et al.}\ }
\newcommand{\via}{{\em via} }
\newcommand{\Sec}{Sec.\ }
\newcommand{\Fig}{Fig.\ }

\newcommand{\Eq}{Eq.\  }
\newcommand{\Eqs}{Eqs.\  }
\newcommand{\rr}{\mbox{$\mathbf{r}$}}
\newcommand{\qq}{\mbox{$\mathbf{q}$}}
\newcommand{\nn}{\mbox{$\mathbf{n}$}}
\newcommand{\NN}{\mbox{$\mathbf{N}$}}
\newcommand{\hq}{\mbox{$\hat{q}$}}
\newcommand{\hkk}{\mbox{$\hat{\kappa}$}}
\newcommand{\hk}{\mbox{$\hat{k}$}}
\newcommand{\QQ}{\mbox{$\mathbf{Q}$}}
\newcommand{\Tr}{\mbox{Tr}}
\newcommand{\bS}{\mbox{$\overline{S}$}}
\begin{document}
\title{Nematic liquid crystals at rough and fluctuating interfaces}
\author{Jens~Elgeti\inst{1}\thanks{\emph{Present address:} 
Institute of Solid State Research, Research Center J\"ulich, D-52425 J\"ulich, Germany}%
\and Friederike~Schmid\inst{1}
}                     
%
%
\institute{Theoretische Physik, Universit\"at Bielefeld, D-33501 Bielefeld, Germany}%
\date{Received: date / Revised version: date}
%
\abstract{
Nematic liquid crystals at rough and fluctuating interfaces are analyzed within 
the Frank elastic theory and the Landau-de Gennes theory. We study specifically
interfaces that locally favor planar anchoring. In the first part we reconsider the
phenomenon of Berreman anchoring on fixed rough surfaces, and derive new simple
expressions for the corresponding azimuthal anchoring energy. Surprisingly, we find 
that for strongly aligning surfaces, it depends only on the geometrical surface 
anisotropy and the bulk elastic constants, and not on the precise values
of the chemical surface parameters.
In the second part, we calculate the capillary waves at nematic-isotropic
interfaces. If one neglects elastic interactions, the capillary wave spectrum
is characterized by an anisotropic interfacial tension. With elastic interactions, 
the interfacial tension, \ie the coefficient of the leading $q^2$ term of 
the capillary wave spectrum, becomes isotropic. However, the elastic interactions 
introduce a strongly anisotropic {\em cubic} $q^3$ term. The amplitudes of 
capillary waves are largest in the direction perpendicular to the director.
These results are in agreement with previous molecular dynamics simulations.
\PACS{
      {61.30 Hn}{Liquid crystals: Surface phenomena} \and  
      {61.30 Dk}{Continuum theories of liquid crystal structure}   \and
      {68.05 -n}{Liquid-liquid interfaces}   
     } 
} 
\maketitle
\section{Introduction}
\label{sec:introduction}

Nematic liquid crystals are fluids with long range orientational order~\cite{degennes}.
Compared to interfaces and surfaces in simple fluids, surfaces of nematic fluids
have several peculiarities. 
First, the interface orients the nematic fluid~\cite{geary,cognard,jerome}.
This phenomenon, called surface anchoring, is quite remarkable, because it implies 
that the surface has direct
influence on a {\em bulk} property of the adjacent fluid. It also has well-known 
practical applications in the LCD (liquid crystal display) 
technology~\cite{raynes}. 
Surface anchoring is driven by energetic and geometric factors, and depends 
on the structure of the surface.
Second, the oriented nematic fluid breaks the planar symmetry of the interface.
This should influence the properties of free interfaces, \eg the spectrum
of capillary wave fluctuations.
Third, the nematic fluid is elastic in a generalized sense, \ie fluctuations of
the local orientation (the director) have long range correlations~\cite{chaikin}.
Since interfacial undulations and director fluctuations are coupled by means of the
surface anchoring, this should introduce long range elastic interactions between the
undulations. Hence interesting effects can be expected from the interplay
of surface undulations and director fluctuations in liquid 
crystals~\cite{rapini1}.

In the present paper, we examine these phenomena in the framework of two continuum 
theories -- the Frank elastic theory and the Landau-de Gennes theory. To separate 
the different aspects of the problem, we first consider a nematic liquid crystal
in contact with a fixed, rough or patterned surface (\Sec \ref{sec:surface}). 
In LCDs, alignment layers are often prepared by coating them with polymers 
(polyimides) and then softly brushing them in the desired direction of 
alignment~\cite{raynes,mauguin,toney}. Assuming that brushing creates grooves 
in the surface~\cite{lee1}, the success of the procedure indicates that liquid crystals 
tend to align in the direction where the surface modulations are smallest. 
Similarly, liquid crystals exposed to surfaces with stripelike gratings were found 
to align parallel to the stripes~\cite{lee2,lee3,rastegar,behdani}. Molecular factors of 
course contribute to this phenomenon, but the effect can already be explained 
on the level of the elastic theory. This was first shown by Berreman~\cite{berreman}, 
and the theory was later refined by different authors~\cite{faetti,fournier1,fournier2}. 
Here we reconsider the phenomenon and derive simple new expressions for
the anchoring angle and the anchoring strength.

In the second part (\Sec \ref{sec:capillary}), we consider the capillary
wave spectrum of free nematic/isotropic interfaces. Capillary waves are
soft mode fluctuations of fluid-fluid interfaces, that are present
even in situations where fluctuations can otherwise bee neglected. 
They were first predicted by Smoluchowski~\cite{smoluchowski}, and the theory was later 
worked out by various authors~\cite{buff,rowlinson,weeks,bedeaux,parry,mecke,stecki}.
Since then they were observed in various systems 
experimentally~\cite{doerr,fradin,mora,li,aarts}
as well as in computer 
simulations~\cite{mon,schmid,werner1,grest,werner2,akino,vink1,mueller,germano}.
In the simplest case, the capillary wave spectrum is governed by the 
free energy cost of the interfacial area that is added due to the undulations. 
Assuming that the fluctuating interface position can be parametrized 
by a single-valued function
$h(x,y)$, and that local distortions $\partial h/\partial x$ and
$\partial h/\partial y$ are small, the thermally averaged squared 
amplitude of fluctuations with wavevector \qq\ is predicted to be
\begin{equation}
\label{eq:cap}
\langle |h(\qq)|^2 \rangle = \frac{k_B T}{\sigma q^2},
\end{equation}
where $\sigma$ is the interfacial tension. Note that 
$\langle | h(\qq) |^2 \rangle$ diverges in the limit $q \to 0$, hence 
the capillary waves with long wavelengths are predicted to be quite
large. In real systems, however, the two coexisting fluids usually 
have different mass densities, and the gravitation introduces
a low-wavelength cutoff in \Eq (\ref{eq:cap}). 

In the last years, capillary waves have attracted renewed interest in the context
of soft condensed matter science. This is mainly due to the fact that typical 
interfacial tensions in soft materials are low, typical length scales are large, 
and coexisting phases often have very similar mass densities. Therefore, the 
capillary wave amplitudes in soft materials tend to be much larger than in simple 
fluids. For example, capillary waves were shown to have a significant effect on 
experimentally measured interfacial widths in polymer 
blends~\cite{sferrazza,klein,carelli}.
Recently, Aarts \etal have even succeeded in visualizing capillary waves directly 
in a colloid-polymer mixture~\cite{aarts}. 

Liquid crystals are a particularly interesting class of soft materials,
because of the additional aspect of orientational order. The present
study is partly motivated by a recent simulation studies of the 
nematic/isotropic interface in a system of ellipsoids~\cite{akino},
where it was found that (i) the capillary
wave spectrum is anisotropic, and (ii) the interface is rougher on short
length scales than one would expect from \Eq (\ref{eq:cap}). While the second
observation is not unusual~\cite{stecki} and has been predicted theoretically for 
systems with short range~\cite{parry} and long range interactions~\cite{mecke}, 
the first is clearly characteristic for liquid crystal interfaces. 
In \Sec \ref{sec:capillary},
we will analyze it within the Landau-de Gennes theory. In particular, we will
discuss the influence of elastic interactions. We find that the anisotropy of 
the spectrum can already been explained within an approximation that excludes 
elastic interactions. However, adding the latter changes the spectrum qualitatively 
to the effect that the leading surface tension term becomes isotropic, and the 
anisotropy is governed by additional higher order terms. 
We summarize and conclude in \Sec \ref{sec:summary}.

\section{Berreman anchoring on rough and patterned surfaces}
\label{sec:surface}

We consider a nematic liquid crystal confined by a surface at $z = h(x,y)$,
which locally favors planar anchoring (\ie alignment parallel to the surface). 
The surface fluctuations $h(x,y)$ are assumed to be small. 
The bulk free energy is given by the Frank elastic energy~\cite{degennes,frank}
\begin{eqnarray}
\nonumber
F_F &=& \frac{1}{2} \int dx \: dy \int_{-\infty}^{h(x,y)} \!\! dz \:
\Big\{ 
K_1 (\nabla \nn)^2 + K_2 (\nn (\nabla \times \nn))^2 
\\ & & 
+ \: K_3 (\nn \times (\nabla \times \nn))^2
\Big\},
\label{eq:frank}
\end{eqnarray}
where $\nn$ is the director, a vector of length unity which describes the local
orientation of the liquid crystal, and $K_i$ are the elastic constants (splay,
twist and bend). Since the surface favors planar alignment and the bulk fluctuations
are small, we
assume that the orientation of the director deep in the bulk, $\nn_b$, 
lies in the $(x,y)$-plane and that local deviations from $\nn_b$ are small.  
Without loss of generality, we take $\nn_b$ to point in the $y$ direction. 
Hence we rewrite the director as
\begin{equation}
\label{eq:director}
\nn = (u, \sqrt{1 - u^2 - v^2}, v),
\end{equation}
and expand the free energy (\ref{eq:frank}) up to second order in powers
of $u$, $v$, and $h$. This gives
\begin{eqnarray}
\nonumber
F_F &\approx& \frac{1}{2} \int dx \: dy \int_{-\infty}^{0} \!\! dz \:
\Big\{
K_1 (\partial_x u + \partial_z v)^2 
\\ &&
+ \: K_2 (\partial_z u - \partial_x v)^2
+ K_3 ( (\partial_y u)^2 + (\partial_y v)^2 ) \Big\}.
\label{eq:frank2}
\end{eqnarray}
Next we perform a two dimensional Fourier transform $(x,y) \to \qq$.
Minimizing $F_F$ in the bulk leads to the Euler-Lagrange equations
\begin{displaymath}
\left( \begin{array}{cc}
K_2 \partial_{zz} \! - \! K_1 q_x^2 \! - \! K_3 q_y^2 
&  - i q_x (K_2\! -\! K_1) \partial_z \\
- i q_x (K_2\! -\! K_1) \partial_z & 
K_1 \partial_{zz} \!-\! K_2 q_x^2\! -\! K_3 q_y^2 
\end{array} \right)
\left( \begin{array}{c} u \\ v \end{array} \right)
= 0.
\end{displaymath}
For the boundary conditions $(u,v) \to 0$ for $z \to - \infty$ and
$(u,v) = (u_0,v_0)$ at $z = 0$, the solution has the form
\begin{equation}
\label{eq:uv}
\left( \begin{array}{c} u \\ v \end{array} \right) = 
\left( \begin{array}{cc} i q_x & - \lambda_2 \\ \lambda_1 & i q_x \end{array} \right)
\left( \begin{array}{c} c_1 \exp(\lambda_1 z) \\ c_2 \exp(\lambda_2 z)
\end{array} \right)
\end{equation}
with the coefficients
\begin{equation}
\label{eq:coeff}
\left( \begin{array}{c} c_1 \\ c_2 \end{array} \right) =
\frac{1}{\lambda_1 \lambda_2 - q_x^2}
\left( \begin{array}{cc} i q_x & \lambda_2 \\ - \lambda_1 & i q_x \end{array} \right)
\left( \begin{array}{c} u_0 \\ v_0 \end{array} \right)
\end{equation}
and the inverse decay lengths
\begin{equation}
\label{eq:lambda}
\lambda_{1,2}^2 = q_x^2 + q_y^2 \frac{K_3}{K_{1,2}}.
\end{equation}
Inserting that into the Frank energy (\ref{eq:frank2}), one obtains
\begin{equation}
\label{eq:frank3}
F_F = \frac{1}{2} \! \int \!\! d \qq 
\frac{q_y^2 K_3}{\lambda_1 \lambda_2 \!-\! q_x^2}
\Big\{ \lambda_1 |u_0|^2\!+ \lambda_2 |v_0|^2\!
+ 2 q_x \Im(v_0^* u_0) \Big\},
\end{equation}
where $u_0(\qq)$ and $v_0(\qq)$ are the values of $u(\qq), v(\qq)$ at the surface. 
This is a general result, which we shall also use in \Sec \ref{sec:capillary}.

Now we study more specifically a liquid crystal in contact with a fixed
patterned surface (fixed $h(x,y)$), which anchors in an unspecifically planar way. 
The surface energy is taken to be of Rapini Papoular type~\cite{rapini2}
\begin{equation}
\label{eq:rapini}
F_s = \sigma_0  \int \!\! dA \: 
(1 + \frac{\alpha_0}{2} (\nn_0 \NN)^2 )
\qquad \mbox{with} \qquad 
\sigma_0 > 0, 
\end{equation}
where $dA = dx \: dy \: \sqrt{1 + (\partial_x h)^2 + (\partial_y h)^2}$
is the local surface area element at $(x,y)$, and
\begin{equation}
\label{eq:normal}
\NN =   \frac{1} {\sqrt{1 + (\partial_x h)^2 + (\partial_y h)^2}} 
\: (- \partial_x h, -\partial_y h,1) 
\end{equation}
the local surface normal. Planar anchoring implies $\alpha_0 > 0$.
As before, we rewrite the director at the surface $\nn_0$
in terms of local deviations $u_0$, $v_0$, according to \Eq (\ref{eq:director}),
perform a Fourier transform $(x,y) \to \qq$, and expand $F_s$ up to second order in 
$u_0, v_0$, and $h$. Omitting the constant contribution $\sigma_0 A$, this gives
\begin{equation}
\label{eq:rapini2}
F_s = \frac{\sigma_0}{2} \! \int \! \! d \qq \:
\Big\{ |h|^2 q^2 + \alpha_0 |v_0 - i q_y h |^2 \Big\}.
\end{equation}
We combine (\ref{eq:frank3}) and (\ref{eq:rapini2}) and minimize the total
free energy $F = F_F + F_s$ with respect to $u_0$ and $v_0$. The result is
\begin{equation}
\label{eq:ftot0}
F = \frac{1}{2} \! \int \!\! d \qq \: |h|^2 q^2\:
\Big\{\sigma_0 + K_3 q
\frac{\hq_y^4 \hkk(\hq_y^2) }{1+q \: \hq_y^2 \hkk(\hq_y^2) \: K_3 /\sigma_0 \alpha_0} \Big\}
\end{equation}
with $\hq_y = q_y/q$ and
\begin{equation}
\label{eq:kappa}
\hkk(\hq_y^2) = 1 /\sqrt{1 + \hq_y^2 (K_3/K_1 - 1)}.
\end{equation}
The result can be generalized easily for the case that the bulk director $\nn_b$ 
points in an arbitrary planar direction
\begin{equation}
\label{eq:director2}
\nn_b = (\cos \phi_0, \sin \phi_0, 0)
\end{equation}
by simply replacing $\hq_y$ with $\nn_b\qq/q$.

\Eq (\ref{eq:ftot0}) already shows that the bulk director will favor
orientations where the amplitudes $|h(\qq)|$ are small, \ie the roughness
is low. To quantify this further, we expand the integrand of (\ref{eq:ftot0})
for small wave vectors in powers of $q$. 
The angle dependent part of the free energy as a function of 
the bulk director angle $\phi_0$ then takes the form
\begin{equation}
F(\phi_0) = \frac{K_3}{2} \! \int_{- \pi}^{\pi} \! \! d \phi \:
\cos^4 (\phi - \phi_0) \: \hkk(\cos^2 (\phi - \phi_0)) \: H(\phi),
\end{equation}
where the roughness spectrum $|h(\qq)|^2$ enters the anchoring energy
solely through the function
\begin{equation}
\label{eq:hphi}
H(\phi) = \int_0^{\infty} \! dq \: q^4 |h(\qq)|^2.
\end{equation}
It is convenient to expand $H(\phi)$ into a Fourier series with coefficients
\begin{equation}
\label{eq:hn}
H_n = \frac{1}{2 \pi} \int \! d \phi \: H(\phi) e^{-i n \phi} =:
- |H_n| e^{i n \alpha_n}.
\end{equation}
Similarly, we define
\begin{equation}
c_n = \frac{1}{2 \pi} \int \! d \phi \: \cos^4 \!\!\phi \:\: \hkk(\cos^2\!\!\phi) 
\: e^{-i n \phi}.
\end{equation}
The coefficients $c_n$ are real and vanish for odd $n$.
In the case $K_3 = K_1$ (\eg in the Landau-de Gennes theory, \Sec~\ref{sec:landau}), 
one has $\hkk \equiv 1 $, and the series $c_n$ stops at $|n| = 4$ with $c_2 = 1/4$ 
and $c_4 = 1/16$. In real materials~\cite{degennes}, the elastic constant $K_3$ is 
typically larger than $K_1$ by a factor of 1-3, and the series does not stop.
However, the coefficients for $|n| \le 4$ remain positive, and the
coefficients for $|n| > 4$ become very small, such that they may be neglected. 

Omitting constant terms that do not depend on $\phi_0$, the anchoring energy 
can then be written as
\begin{equation}
\label{eq:anchoring_energy}
F(\phi_0) = - \pi K_3 \sum_{n = 2,4} c_n |H_n| \cos n(\phi_0 - \alpha_n).
\end{equation}
The anchoring angle is the angle that minimizes $F(\phi_0)$. In general,
the $n=2$ term will dominate, and one gets approximately
\begin{equation}
\label{eq:anchoring_angle}
\bar{\phi}_0 \approx \alpha_2 =
\frac{1}{2} \mbox{arg}\Big( - \frac{1}{2 \pi} \int d\phi \: e^{-2 i \phi} H(\phi)\Big).
\end{equation}
We note that the angles $\alpha_n$ in \Eq (\ref{eq:hn}) correspond to directions 
where the height fluctuations are small, because the contributions of $H_0$ and $H_n$ 
to the spectral function $H(\phi)$ have opposite signs. Hence \Eq (\ref{eq:anchoring_angle}) 
implies that the surface aligns the nematic fluid in a direction where the surface
is smooth.  At given anchoring angle $\bar{\phi}_0$, 
we can also calculate the anchoring strength. To this end, we expand the anchoring 
energy about $\bar{\phi}_0$ and obtain 
$F(\phi_0) = F(\bar{\phi}_0) + \frac{W}{2} (\phi_0 - \bar{\Phi}_0)^2$ 
with the anchoring strength
\begin{equation}
\label{eq:anchoring_strength}
W = \pi K_3 \sum_{n = 2,4} n^2 \: c_n |H_n| \cos n (\bar{\phi}_0 - \alpha_n).
\end{equation}

We conclude that elastic interactions in nematic liquid crystals on
anisotropically rough surfaces induce an anchoring energy in a direction of 
low roughness. The central quantities characterizing the surface roughness
are the two coefficients $H_{2,4}$ defined by \Eqs (\ref{eq:hphi}) and (\ref{eq:hn}). 
These quantities determine the anchoring strength (\Eq (\ref{eq:anchoring_strength})), 
and the anchoring angle (\Eq (\ref{eq:anchoring_angle})). The anchoring 
mechanism only requires an unspecific tendency of the liquid crystal to 
align parallel to the interface (\Eq (\ref{eq:rapini})). Given such a
tendency, the anchoring energy no longer depends on the surface parameters, 
$\alpha_0$ and $\sigma_0$. The only relevant material parameters are the 
splay and bend elastic constants in the bulk, $K_1$ and $K_3$, and the 
squared surface anisotropy, which is characterized by the coefficients $H_{2,4}$. 

We note that our treatment premises that the nematic liquid stays perfectly
ordered at the surface. In reality, rough surfaces may reduce the 
order, which in turn influences the anchoring properties~\cite{papanek,cheung}. 
This has not been considered here. 

\section{Capillary waves at the nematic/isotropic interface}
\label{sec:capillary}

In this section we study the capillary wave spectrum of freely undulating
nematic/isotropic (NI) interfaces. The problem is similar to that considered
in the previous section (\Sec \ref{sec:surface}), with two differences:
(i) The interface position $h(x,y)$ is free and subject to thermal fluctuations, 
and (ii) the nematic order at the interface drops smoothly to zero. 
The second point implies, among other, that the elastic constants are 
reduced in the vicinity of the interface.

In many systems, the anchoring at NI-interfaces is planar.
As a zeroth order approach, we neglect the softness of the profile and
approximate the interfacial structure by a steplike structure
(sharp-kink approximation), and the interfacial free 
energy by \Eq (\ref{eq:ftot0}) with effective parameters
$\sigma_0$ and $\alpha_0$. Generally the capillary waves of an interface
with an interfacial free energy of the form
\begin{equation}
\label{eq:ftot_sample}
F = \frac{1}{2} \int d \qq \: |h(\qq)|^2 \Sigma(\qq)
\end{equation}
are distributed according to
\begin{equation}
\label{eq:cap_sample}
\langle | h(\qq) |^2 \rangle = k_B T/\Sigma(\qq)).
\end{equation}
Thus the free energy (\ref{eq:ftot0}) yields the capillary wave spectrum
\begin{equation}
\frac{k_B T/\sigma_0}{\langle | h(\qq) |^2 \rangle}  \approx 
q^2 + q^3 \; \frac{K_3}{\sigma_0} \hq_y^4 \hkk
-  q^4 \: \frac{K_3^2 \hq_y^6 \hkk^2}{\sigma_0^2 \alpha_0} + \cdots
\label{eq:cap0_exp}
\end{equation}
As before, $\hq_y$ is the component of the unit vector $\qq/q$ in the
direction of the bulk director.

The result (\ref{eq:cap0_exp}) shows already the three remarkable features, which 
will turn out to be characteristic for the NI interface. First, the capillary 
wave spectrum is anisotropic, the capillary waves in the direction 
parallel to the director ($\hq_y$) being smaller than in the direction perpendicular 
to the director. Second, the leading (quadratic) term is still isotropic; 
the anisotropy enters through the higher order terms. Third, in contrast to 
simple fluids with short range interactions, the capillary wave spectrum cannot 
be expanded in even powers of $q$, but it contains additional cubic 
(and higher order odd) terms. This implies that the capillary wave spectrum 
is nonanalytic in the limit $\qq \to 0$.

These findings are gratifying. However, the sharp kink description of the 
NI interface is inadequate. The Frank free energy (\ref{eq:frank}) describes 
nematic liquid crystals with constant local order parameter, whereas at NI
interfaces, the nematic order parameter drops softly to zero. Moreover, the
surface anchoring at NI interfaces is an intrinsic property of the interface, 
which depends itself on the local elastic constants. We will now consider our 
problem within a unified theory for nematic and isotropic liquid crystals, the 
Landau-de Gennes theory.

\subsection{Landau-de Gennes theory}
\label{sec:landau}

The Landau-de Gennes theory is based on a free energy expansion in powers of
a symmetric and traceless ($ 3 \times 3$) order tensor field $\QQ(\rr)$.
\begin{eqnarray}
\nonumber
F &=& \int \!\!\! d\rr \Big\{ \frac{A}{2} \Tr(\QQ^2) \! + \! \frac{B}{3} \Tr(\QQ^3)
\! + \! \frac{C_1}{4} \Tr(\QQ^2)^2 \! + \! \frac{C_2}{4} \Tr(\QQ^4) 
\\
&&  \qquad
+\: \frac{L_1}{2} \partial_i Q_{jk} \partial_i Q_{jk}
+ \frac{L_2}{2} \partial_i Q_{ij} \partial_k Q_{kj} \Big\}
\label{eq:landau_q}
\end{eqnarray}
Following a common assumption, we neglect the possibility of biaxiality
and rewrite the order tensor as~\cite{priestley}
\begin{equation}
\label{eq:order}
Q_{ij}(\rr) = \frac{1}{2} S(\rr) ( 3 n_i(\rr) n_j(\rr) - \delta_{ij}).
\end{equation}
Here $S$ is the local order parameter, and $\nn$ a unit vector characterizing
the local director. In the homogeneous case ($\partial_i Q_{jk} \equiv 0$),
the free energy (\ref{eq:landau_q}) predicts a first order transition
between an isotropic phase (I) with $S = 0$ and an oriented nematic phase
(N) with $S = S_0 = -2/9 B/(C_1 + C_2/2)$. We recall briefly the 
properties of a flat NI interface at coexistence for a system with fixed 
director $\nn$, as obtained from minimizing (\ref{eq:landau_q})~\cite{priestley}: 
The order parameter profile has a simple tanh form
\begin{equation}
\label{eq:profile}
S(z) = S_0 \bS(z/\xi) 
\qquad \mbox{with} \quad
\bS(\tau) = 1/(e^{\tau} + 1).
\end{equation}
The interfacial width
\begin{equation}
\label{eq:width}
\xi = \xi_0 \sqrt{1 + \alpha (\nn \NN)^2}
\quad \Big( \xi_0 = \frac{2}{S_0} 
\sqrt{\frac{L_1 + L_2/6}{3 (C_1 + C_2/2)}} \Big)
\end{equation}
and the interfacial tension
\begin{equation}
\label{eq:tension}
\sigma = \sigma_0 \sqrt{1 + \alpha (\nn \NN)^2}
\quad \Big(\sigma_0 = \frac{3(C_1 + C_2/2)}{16} 
S_0^4 \xi_0 \Big)
\end{equation}
both depend in the same way on the angle between the director $\nn$ 
and the surface normal $\NN$, \via the parameter 
\begin{equation}
\alpha = \frac{1}{2} \: \frac{L_2}{L_1 + L_2/6}.
\end{equation}

The quantity $\sigma_0$ sets the energy scale, $\xi_0$ the length scale, 
and $S_0$ the ``order parameter scale''. (Note that $S$ can be rescaled even
though it is dimensionless). Hence only one characteristic material parameter
remains, \eg the parameter $\alpha$. In the following, we shall always 
use rescaled quantities $S \to S/S_0$, $\mbox{length} \to \mbox{length}/\xi_0$, 
$\mbox{energy} \to \mbox{energy}/\sigma_0$. The free energy at 
coexistence can then be rewritten as~\cite{priestley} 
\begin{equation}
\label{eq:landau}
F = \int d \rr \: \{ f + g_1 + g_2 + g_3 + g_4 \}
\end{equation}
\begin{eqnarray*}
\mbox{with} \quad
f &=& 3 S^2 (S^2 - 1)  \qquad \mbox{(at coexistence)}\\
g_1 &=& 3 \Big( (\nabla S)^2 + \alpha (\nn \nabla S)^2 \Big) \\
g_2 &=& 12 \alpha \Big( (\nabla \nn) (\nn \nabla S)
+ \frac{1}{2} (\nn \times \nabla \times \nn) (\nabla S) \Big)\\
g_3 &=& 3 \Big(
(3 + 2 \alpha) (\nabla \nn)^2 
+ (3 - \alpha) (\nn \cdot \nabla \times \nn)^2 \\
&& + \: (3 + 2 \alpha)(\nn \times \nabla \times \nn)^2
\Big).
\end{eqnarray*}
The first term $f(S)$ describes the bulk coexistence, the middle terms
$g_1$ and $g_2$ determine the structure of the interface, and the
last term establishes the relation to the Frank elastic energy,
\Eq (\ref{eq:frank}). We note that in this version of the Landau-de Gennes 
theory, the splay and the bend elastic constants are identical, $K_1=K_3$,
hence $\hkk(\hq_y^2) \equiv 1 $ in \Eq (\ref{eq:cap0_exp}).

\Eq (\ref{eq:landau}) will be our starting point.
As in \Sec \ref{sec:surface}, we will assume without loss of generality 
that the interface is on average located at $z=0$, and that the bulk director 
far from the surface points in the $y$-direction.

\subsection{Constant director approximation}
\label{sec:constant}

We return to considering undulating interfaces with varying position $h(x,y)$.
In the simplest approach, the director is constant throughout the system, 
$\nn \equiv (0,1,0)$. Elastic interactions are thus disregarded.
For the order parameter, we make the Ansatz $S(\rr) = \bS((z - h(x,y))/\xi)$, 
where $\bS$ is the tanh profile from \Eq (\ref{eq:profile}), and the 
interfacial width $\xi$ varies with the local surface normal 
$\NN$ (\ref{eq:normal}) according to \Eq (\ref{eq:width}).  
Inserting this into the free energy (\ref{eq:landau}), 
and Fourier transforming $(x,y) \to \qq$, one obtains
\begin{equation}
\label{eq:ftot_fixed}
F = A + \frac{1}{2} \int \! d\qq \: |h|^2 \Big\{ q^2 + \alpha q_y^2 \Big\}.
\end{equation}
This result is quite robust. As we shall see in \Sec (\ref{sec:local}), it
is also obtained with a rather different Ansatz for $S$, as long
as the director is kept constant. We can now apply \Eq (\ref{eq:cap_sample}) 
and obtain the capillary wave spectrum
\begin{equation}
\label{eq:cap_fixed}
\frac{k_B T/\sigma_0}{\langle | h(\qq) |^2 \rangle} = (q^2 + \alpha q_y^2)
\end{equation}
(recalling that wave vectors $q$ are given in units of $1/\xi_0$ (\ref{eq:width})),
which is anisotropic. 

Hence already this simple approach predicts anisotropic capillary wave amplitudes.
The capillary waves are weakest in the $y$ direction, which is the direction of the 
bulk director. It is interesting that we get the anisotropy already at this point. 
One might have suspected that the dampening of waves parallel to the director is
caused by director-director interactions. This turns out not to be the case, 
instead an interaction between the director and the order parameter gradient 
$\nn \nabla S$ is responsible for the anisotropy. As the director wants to
align parallel to the surface, waves parallel to the director have higher energy.

\subsection{Relaxing the director: A variational approach}
\label{sec:tilted}
%

Since the interface locally favors parallel anchoring, one would expect 
that the director follows the undulations of the interface (\Fig \ref{fig:surface}). 
This motivates a variational Ansatz $\nn = (0, \cos \theta, \sin \theta)$ with
\begin{equation}
\label{eq:theta}
\theta(x,y,z) = g(x,y) \: \exp[\frac{\kappa}{\xi} (z - h(x,y))].
\end{equation}
As before, we assume that the profile has the form $S = \bS( (z - h)/\xi)$. 
After inserting this Ansatz in (\ref{eq:landau}), expanding in $\theta$ up 
to second order, and minimizing with respect to $g$, a lengthy calculation
yields the surface free energy (omitting constant terms) 
\begin{eqnarray}
\label{eq:ftot_variable}
F &=& \frac{1}{2} \int \!\! d \qq \: |h|^2 
\Big\{ q^2 + \alpha q_y^2(1\! - \!\frac{3 c^2}{c_1 + c_2 q_y^2 + c_3 q_x^2}) \Big\}
\end{eqnarray}
\begin{eqnarray}
\lefteqn{
\mbox{with} \qquad 
c = 
\Big[ {3 \choose 1+ \kappa}^{-1}_{\Gamma} 
- 2 \kappa {2 \choose \kappa}^{-1}_{\Gamma} \Big]
} \nonumber \\
c_1 &=&
\Big[ {3 \choose 1+ 2\kappa}^{-1}_{\Gamma} 
- 2 \kappa {2 \choose 2 \kappa}^{-1}_{\Gamma} 
+ \kappa^2 \frac{3 + \alpha}{\alpha} {1 \choose \kappa-1}^{-1}_{\Gamma} 
\Big] 
\nonumber \\ \nonumber
c_2 &=&
\frac{3 + 2 \alpha}{3 \alpha} 
{1 \choose 2 \kappa-1 }^{-1}_{\Gamma} ,
\qquad
c_3 = 
\frac{3 - \alpha}{\alpha},
{1 \choose 2 \kappa-1}^{-1}_{\Gamma} 
\end{eqnarray}
where we have defined generalized binomial coefficients,
\begin{displaymath}
{n \choose a}_{\Gamma} = \frac{\Gamma(n+1)} {\Gamma(a+1) \Gamma(n-a+1)} .
\end{displaymath}
As a consistency check, we also inspected the result for $\theta(x,y,z)$ 
directly. It is proportional to $\partial_y h$ as expected.

The comparison of the free energy (\ref{eq:ftot_variable}) with the 
corresponding result for fixed director, \Eq (\ref{eq:ftot_fixed}), shows 
that the anisotropy of the surface fluctuations is reduced. 
For a further analysis, it would be necessary to minimize the free energy
expression (\ref{eq:ftot_variable}) with respect to the variational
parameter $\kappa$. Unfortunately, $\kappa$ enters in such a complicated
way, that this turns out to be unfeasible. 
Numerically, we find that the capillary wave spectrum, obtained
\via \Eq (\ref{eq:cap_sample}), varies only little with $\kappa$.
For any reasonable value of $\kappa$, \ie $\kappa^{-1}>2$,
the result differs from that obtained with the constant director 
approximation (\Eq (\ref{eq:cap_fixed})) by less than one percent. 
Within the present approximation, the effect of relaxing the director 
is negligeable. This is mostly due to the fact that \Eq (\ref{eq:theta}) 
still imposes rather rigid constraints on the director variations in 
the nematic fluid.

\subsection{Local profile approximation}
\label{sec:local}
A more general solution can be obtained with the additional approximation
that the width of the interface is small, compared to the relevant length
scales of the interfacial undulations. In that case, the interface and the bulk 
can be considered separately, and we can derive analytical expressions 
for the capillary wave spectrum.
The assumption that length scales can be separated is highly questionable,
because interfacial undulations are present on all length scales down 
to the molecular size. Nevertheless, computer simulations of other 
systems (Ising models~\cite{mueller} and polymer blends~\cite{werner2})
have shown that intrinsic profile models can often describe the
structure of interfaces quite successfully. 

\begin{figure}[htbp]
\centerline{
\resizebox{0.3\textwidth}{!}{ 
\includegraphics{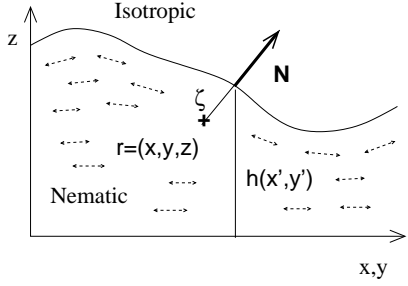}}
}
\caption{Nematic-Isotropic interface with local coordinates.
\label{fig:surface}
}
\end{figure}

We separate the free energy (\ref{eq:landau}) into an interface and
bulk contribution, $F = F_S + F_F$. The bulk contribution $F_F$ 
has the form (\ref{eq:frank}) with the elastic constants
$K_1 = K_3 = 6(3+2 \alpha)$ and $K_2 = 6 (3-\alpha)$, and accounts
for the elastic energy within the nematic region. The integrand
in the expression for the remaining surface free energy $F_S$ 
vanishes far from the surface. We assume that the local order 
parameter profile has mean-field shape in the direction 
perpendicular to the surface.  
More precisely, we make the Ansatz
\begin{equation}
\label{eq:ss_appr}
S(\rr) = \bS (\zeta/\xi), \qquad
\nabla S \approx \frac{1}{\xi} 
(\frac{d \bS}{d \tau}\Big|_{\tau = \zeta/\xi} \NN,
\end{equation}
(cf. (\ref{eq:profile}), (\ref{eq:width})), where $\NN$ is
the local surface normal as usual, and $\zeta$ is the 
distance between $\rr$ and the closest interface point.
The $(x,y)$ coordinates at this point are denoted $(x',y')$.
(see \Fig \ref{fig:surface}). To evaluate $F_S$, we make a coordinate 
transformation $\rr \to (x',y', \zeta)$ and integrate over $\zeta$.
The relation between the coordinates is $\rr = (x',y',h(x',y')) + \NN \zeta$, 
and the Jacobi determinant for the integral is in second order 
of $h$ 
\begin{displaymath}
1 + \frac{1}{2}((\partial_x h)^2 + (\partial_y h)^2)
- \zeta (\partial_{xx} h + \partial_{yy} h)
+ \zeta^2 (h_{xx} h_{yy} - h_{xy} h_{yx}).
\end{displaymath}

We begin with reconsidering the constant director case, $\nn = \mbox{const}$. 
The Frank free energy then vanishes, and the surface free energy takes 
exactly the form of \Eq (\ref{eq:ftot_fixed}). Hence the present
approximation leads to the same expression as the approximation 
taken in \Sec \ref{sec:constant}. This underlines the robustness of
the result (\ref{eq:cap_fixed}).

In the general case, we must make an Ansatz for the variation of
the director $\nn$ in the vicinity of the surface. We assume
that it varies sufficiently slowly, so that we can make a linear
approximation
\begin{equation}
\label{eq:nn_linear}
\nn(x',y',\zeta) \approx
\nn(x',y',0) + \zeta (\NN \cdot \nabla) \nn.
\end{equation}
As before, we take the bulk director to point
in the $y$-direction. The local director deviations in 
the $x$ and $z$ direction are parametrized in terms of two 
parameters $u$ and $v$ according to \Eq (\ref{eq:director}). 

After inserting \Eqs (\ref{eq:ss_appr}) and (\ref{eq:nn_linear})
into \Eq (\ref{eq:landau}), expanding up to second order in 
$h$, $u$, and $v$, and some partial integrations, we obtain 
the surface free energy
\begin{eqnarray}
\label{eq:fs}
F_S &=& \int dx \: dy \: (g_{1s} + g_{2s} + g_{3s})
\end{eqnarray}
with
\begin{eqnarray}
g_{1s} &=& 1 + \frac{1}{2} ((\partial_x h)^2 + (\partial_y h)^2)
\nonumber \\&& \quad \nonumber
+ \frac{\alpha}{2} (\partial_y h - v_0)^2
+ \frac{\pi^2 - 6}{6} \alpha (\partial_z v_0)^2
\nonumber \\ \nonumber
g_{2s} &=& 
3 \alpha \Big( 
   2 (\partial_y h)(\partial_x u_0 + \partial_z v_0) 
   + (\partial_{xx} h + \partial_{yy} h) (\partial_y v_0)
\nonumber \\&& \nonumber
   - (\partial_x h) (\partial_y u_0) 
   - v_0 (\partial_z u_0 - 3 \partial_x u_0)
\nonumber \\&& \nonumber
   + 2 (\partial_z v_0) (\partial_x u_0 + \partial_z v_0)
   + 2 (\partial_z u_0) (\partial_z u_0 - \partial_x v_0)
   \Big)
\nonumber \\ \nonumber
\nonumber
g_{3s} &=& 
3 \Big( (3 \!+\! 2 \alpha) (\partial_x u_0 \!+\! \partial_z v_0)^2 
+ (3 \!- \! \alpha) (\partial_z u_0 \!- \!\partial_x v_0)^2
\\ \nonumber &&
\quad
+ \: (3 \!+\! 2 \alpha)( (\partial_y u_0)^2 \!+\! (\partial_y v_0)^2)
\Big),
\nonumber 
\end{eqnarray}
where $u_0$ and $v_0$ are the values of $u$ and $v$ at the interface.
The first contribution $g_{1s}$ describes the effect of the 
anisotropic local surface tension. The second contribution $g_{2s}$ 
arises from the coupling between the director variations 
$\partial_i n$ with the order parameter variation  $\nabla S$ at the
interface. The last contribution $g_{3s}$ accounts for the reduction 
of the Frank free energy in the interface region.

The ensuing procedure is similar to that of \Sec \ref{sec:surface}. 
We first minimize the bulk free energy $F_F$, which leads in the 
most general case to \Eq (\ref{eq:ftot0}). Then we minimize the 
total energy $F = F_F + F_S$ with respect to $u_0$ and $v_0$,
using \Eqs (\ref{eq:uv}) and (\ref{eq:coeff}) to estimate
the derivatives $\partial_z u_0$, $\partial_z v_0$.
The result has the form (\ref{eq:ftot_sample}) and gives
the capillary wave spectrum \via (\ref{eq:cap_sample}). 
Unfortunately, the final expression is a rather lengthy, 
and we cannot give the formula here. We will discuss it
further below. 

A more concise and qualitatively similar result is obtained with 
the additional approximation, that the director only varies in the 
$z$ direction, $u \equiv 0$. In that case, the minimization 
of the Frank free energy (\ref{eq:frank}) with respect to $v$ yields
$v(\qq,z) = v_0 \exp(q \hk z)$ with
\begin{equation}
\hk(\hq_y^2) = \sqrt{\frac{3 - \alpha}{3 + 2 \alpha}}
\sqrt{1 + \hq_y^2 \frac{3 \alpha}{3 - \alpha}} 
\end{equation}
($\hq_y = q_y/q$), \ie at $z=0$ we have $\partial_z u_0 = q \hk u_0$. 
The Frank energy takes the form
\begin{equation}
F_F = \frac{1}{12 \alpha} \int d\qq \:
(3 + 2 \alpha) \: q \hk(\hq_y^2) \: |v_0|^2.
\end{equation}
This equation replaces \Eq (\ref{eq:ftot0}). The surface
free energy $F_S$ (\ref{eq:fs}) is also greatly simplified. 
After minimizing the sum $F=F_F + F_S$ with respect to $v_0$ 
and applying \Eq (\ref{eq:cap_sample}), one obtains the 
capillary wave spectrum
\begin{eqnarray}
\lefteqn{
\frac{k_B T/\sigma_0}{\langle | h(\qq) |^2 \rangle} 
 =  q^2 + \alpha q_y^2 \: 
} 
\nonumber \\
\label{eq:cap_local1}
&&
\times \Big( 1  -
 \frac{(1 - 6 q \hk - 3 q^2)^2}{1 + 18 q \hk
((\frac{1}{3}+\frac{1}{\alpha})(1-2 q \hk) 
+ \frac{\pi^2 - 6}{54} q \hk)}
\Big).
\end{eqnarray}
Expanding this for small wavevectors $q$ gives
\begin{equation}
\label{eq:cap_local1_exp}
\frac{k_B T/\sigma_0}{\langle | h(\qq) |^2 \rangle} 
 = 
q^2 + \hq_y^2 ( q^3  \; 18 (1+ \alpha) \hk 
- q^4 \cdots ),
\end{equation}
where the coefficient of the fourth order term is negative. 

Comparing this solution to \Eq (\ref{eq:cap_fixed}), one notes obvious 
qualitative differences. In the constant director approximation,
\Eq (\ref{eq:cap_fixed}), one has an anisotropic effective surface 
tension: The capillary wave spectrum has the form (\ref{eq:cap}) 
with $\sigma = \sigma_0 (1+ \alpha \hq_y^2)$. The present 
treatment shows that the elastic interactions remove the 
anisotropy in the surface tension term (order $q^2$),
but introduce new anisotropic terms that are of higher order
in $q$. This is consistent with the preliminary results
from our earlier zeroth order approach, \Eq (\ref{eq:cap0_exp}). 

\begin{figure}
\centerline{
\resizebox{0.4\textwidth}{!}{ 
\includegraphics{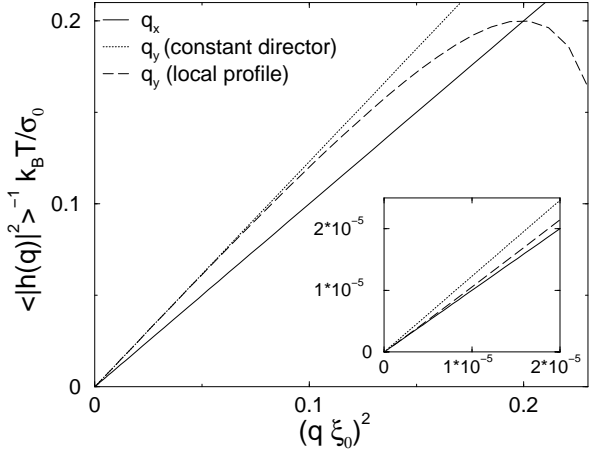}}
}
\vspace*{0.3cm}
\caption{
Capillary wave spectrum in the direction perpendicular to the
bulk director $(q_x)$ (solid line) and parallel to the bulk director 
($q_y$) as obtained from the constant director approximation 
(dotted line) and the local profile approximation (dashed line). 
In the perpendicular direction, both approaches give the same result.
Note the presence of an unphysical pole at $(q \xi_0)^2 = 0.24$. 
The inset shows a blowup for very small $q$-vectors,
illustrating that the spectrum becomes isotropic in the
limit $q \to 0$. The parameter is $\alpha = 3/13$, corresponding
to $L_1 = 2 L_2$.
\label{fig:local1}}
\end{figure}

We turn to discussing the full solution of the local profile
approximation, where both variations of $u$ and $v$ are
allowed. In the directions parallel and perpendicular to the bulk
director ($x$ and $y$), the capillary wave spectrum turns
out to be the same as in (\ref{eq:cap_local1}). It is shown
in \Fig \ref{fig:local1} for a typical value of $\alpha$ 
(taken from Ref.~\cite{priestley}) and compared to the constant director
approximation. The capillary waves in the direction perpendicular 
to the bulk director (the $x$-direction) are not influenced 
by the elastic interactions: The amplitudes only contain a 
$q^2$ contribution and are identical in the constant director
approximation and the local profile approximation. In contrast,
the capillary waves in the direction parallel to the bulk 
director (the $y$-direction) are to a large extent dominated
by the cubic term. The fact that the spectrum becomes isotropic
in the limit $q \to 0$ becomes only apparent at very small
$q$ vectors, $q \xi_0 < 0.005$ (see inset of \Fig \ref{fig:local1}). 

\begin{figure}[htbp]
\resizebox{0.5\textwidth}{!}{ 
\includegraphics{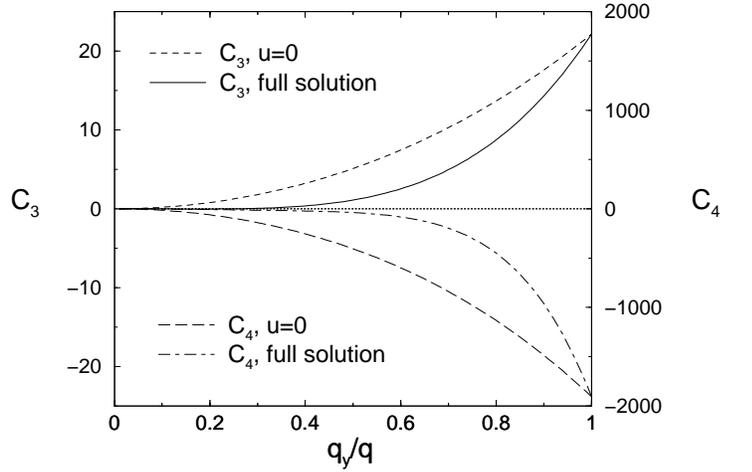}}
\caption{Coefficients of (a) the cubic term $q^3$ 
and (b) the fourth order term $q^4$ vs. $\hq_y = q_y/q$
in the expansion (\protect\ref{eq:cap_local2_exp}) and
(\protect\ref{eq:cap_local1_exp}), at $\alpha = 3/13$.
\label{fig:coeff}}
\end{figure}

The effect of relaxing both $u$ and $v$ in \Eq (\ref{eq:director})
becomes apparent when examining directions that are intermediate
between $x$ and $y$. The expansion of the full solution in powers of
$q$ gives an expression that is similar to \Eq (\ref{eq:cap_local1_exp}),
\begin{equation}
\label{eq:cap_local2_exp}
\frac{k_B T/\sigma_0}{\langle | h(\qq) |^2 \rangle} 
 = 
q^2 + \hq_y^2 ( q^3  \: C_3(\hq_y^2) + q^4 \: C_4(\hq_y^2) + \cdots)
\end{equation}
but has different coefficients $C_i$. \Fig \ref{fig:coeff} shows the coefficients 
$C_3$ and $C_4$ as a function of $\hq_y$ and compares them with the corresponding 
quantities obtained from the simplified solution (\ref{eq:cap_local1}). 
In the full solution, the $C_i$ stay much smaller in the vicinity of 
$q_y \sim 0$. This demonstrates once more that elastic interactions reduce the 
anisotropy of the capillary waves.

Unfortunately, we also find that the free energy 
$F = F_F + F_S$ (from \Eqs (\ref{eq:ftot0}) and (\ref{eq:fs})),
as a functional of $u_0(\qq), v_0(\qq)$, becomes 
unstable for larger wavevectors $q$. The instability
gives rise to the unphysical pole at $(q \xi_0)^2 = 0.24$
in \Fig \ref{fig:local1}. 
In the direction $\hq_y = 1$, the pole is encountered at 
$q \xi_0 \sim 0.5$ for all $\alpha$. In other directions, 
it moves to even smaller $q$ values, which reduces the 
stability further. At $\alpha = 3/13$, the free energy is stable 
only up to $q \xi_0 \sim 0.023$, corresponding to a length scale of 
$\sim 270 \xi_0$. The approximation is bound to break 
down on smaller length scales.

\begin{figure}[htbp]
\resizebox{0.4\textwidth}{!}{ 
\includegraphics{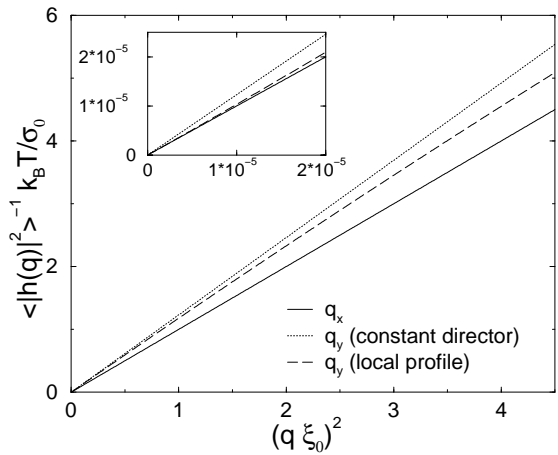}}
\vspace*{0.3cm}
\caption{ Capillary wave spectrum from the local profile
approximation with independent surface derivative
$\partial_z u$, compared to the constant director
approximation, in the directions parallel and
perpendicular to the bulk director. The pole in
\Fig \protect\ref{fig:local1} has disappeared. 
At small wavevectors, the curves are the same
as in the approximation of \Fig \protect\ref{fig:local1}
\label{fig:local3}}
\end{figure}

Hence the region of validity of the theory is very small. 
The instability is presumably caused by the way the local profile 
approximation was implemented. In particular, our estimate for 
the values of the partial derivatives $\partial_z u_0$ and 
$\partial_z v_0$ at the interface must be questioned. They 
were obtained from extrapolating the bulk solution, which is 
however only valid for constant, saturated, order parameter 
$S \equiv S_0$. In order to assess the effect of this constraint, 
we have considered a second approximation: The derivative 
$\partial_z \nn$ at the interface is taken to be independent of 
the bulk solution. In the interface region, the director 
is assumed to vary linearly with the slope $\partial_z \nn$.
At the distance $\sim 2.5 \xi_0$ from the interface,
the profile changes continuously to the exponential bulk 
solution. In this approximation, the value of the derivative 
$\partial_z \nn$ at the surface is an additional variable, 
which can be optimized independently. We have minimized
the free energy with this Ansatz and $u \equiv 0$ 
(\ie no director variations in the $x$-direction),
and calculated the capillary wave spectrum. 
The result is shown in \Fig \ref{fig:local3}. 
The instability from \Fig \ref{fig:local1} disappears.
The other characteristic features of the spectrum remain.
It becomes isotropic for very small wavevectors, 
$ q < 0.003 \xi_0$, corresponding to length scales of 
several thousand correlation lengths $\xi_0$.
On smaller length scales, it is anisotropic. 
It is worth noting that at least on this level of 
approximation, the capillary waves in the direction 
perpendicular to the director (the $x$ direction) 
are not affected by the elastic interactions.

\bigskip

\section{Summary and discussion}
\label{sec:summary}

To summarize, we have studied the interplay of elastic interactions 
and surface undulations for nematic liquid crystals at rough and
fluctuating interfaces using appropriate continuum theories:
the Frank elastic energy and the Landau-de Gennes theory.

In the first part, we have considered nematic liquid crystals
in contact with a surface of given geometry, characterized
by a fixed height function $h(x,y)$. We have re-analyzed
the effect of Berreman anchoring, \ie the phenomenon that
elastic interactions are sufficient to align liquid crystals 
exposed to anisotropically rough surfaces. Our treatment
allowed to derive explicit equations for the anchoring 
angle and the anchoring strength for given (arbitrary)
height function $h(x,y)$. In particular, we find that
the resulting azimuthal anchoring coefficient depends
only on the surface anisotropy and the bulk elastic 
constants $K_1$ and $K_3$, and not on the chemical
surface interaction parameters such as the interfacial
tension and the zenithal anchoring strength.
The contribution of the surface anisotropy to the anchoring 
energy has recently been verified by Kumar et al~\cite{kumar}. 
We hope that our results will stimulate systematic experimental 
research on the role of the elastic constants as well. 

In the second part, we have examined the inverse problem,
the effect of the nematic order on capillary wave
fluctuations of NI interfaces. The work was motivated 
by a previous simulation study, where it was found
that the capillary wave amplitudes were different in
the direction parallel and perpendicular to the bulk director.
Our analysis shows that this effect can be understood
within the Landau-de Gennes theory. As in the simulation,
the waves parallel to the director are smaller than those
perpendicular. The anisotropy is caused by a coupling 
term between the director and the order parameter gradient 
($\nn \nabla S$), which locally encourages the director to
align parallel to the surface, and thus penalizes interfacial 
undulations in the direction of the director.

The influence of elastic interactions mediated by
the nematic bulk fluid was investigated separately.
We find that they reduce the anisotropy and change the 
capillary wave spectrum qualitatively. In the absence of 
elastic interactions, \ie with fixed director, the anisotropy
manifests itself in an anisotropic surface tension. 
If one allows for director modulations, the
surface tension becomes isotropic, and the anisotropy
is incorporated in higher order terms in the
wave vector $q$. In particular, we obtain a large
anisotropic {\em cubic} term. The fourth order term 
is generally negative, \ie we have a negative 
``bending rigidity''. This is consistent with previous 
observations from simulations~\cite{akino}. 
As we have noted in the introduction, bending rigidities 
are often negative in fluid-fluid interfaces, for various 
reasons. In the case of nematic/isotropic interfaces, 
the elasticity of the adjacent medium provides an additional
reason.

We have shown that the higher order terms dominate
the capillary wave spectrum on length scales
up to several thousand correlation lengths. 
This has serious practical consequences.
The analysis of capillary waves is usually
a valuable tool to determine interfacial tensions
from computer simulations. In liquid crystals,
however, this method must be applied with caution.
More generally, our result suggests that the apparent
interfacial tension of nematic/isotropic interfaces should 
be strongly affected by finite size effects. 
In fact, Vink and Schilling have recently reported
that the interfacial tension obtained from computer 
simulations of soft spherocylinders varies considerably 
with the system size~\cite{vink2,vink3}.

We thank Marcus M\"uller for a useful comment, and
the German Science Foundation for partial support.

\end{document}